\documentclass[12pt,titlepage]{amsart}

\usepackage{amssymb,amsfonts,amsmath,amsthm}
\usepackage[makeroom]{cancel}
\usepackage{natbib}
\usepackage{graphicx}

\usepackage{comment}
\usepackage{enumerate}
\usepackage{tikz}\usepgflibrary{shapes}
\usepackage{verbatim}
\usepackage{subfigure}

\def\X{\mathcal{X}}

\def\U{\mathcal{U}}

\def\Re{\mathbf{R}}

\def\argmax{\mbox{argmax}}

\def\la{\lambda}

\def\Rchris{\mathrel{R^C}}
\def\RchrisS{\mathrel{P^C}}
\def\RPref{\mathrel{R^P}}
\def\RPrefS{\mathrel{P^P}}

\newcommand{\df}[1]{\textit{#1}}

\newcommand{\abs}[1]{ \left | #1 \right | }

\newdimen\slantmathcorr
\def\oversl#1{%assuming that mathslant=0.25
\setbox0=\hbox{$#1$}
\slantmathcorr=\wd0
\hskip 0.2\slantmathcorr \overline{\hbox to 0.8\wd0{%
\vphantom{\hbox{$#1$}}}}
\hskip-\wd0\hbox{$#1$}
}

\def\undersl#1{%assuming that mathslant=0.25
\setbox0=\hbox{$#1$}
\slantmathcorr=\wd0
\underline{\hbox to 0.8\wd0{%
\vphantom{\hbox{$#1$}}}}
\hskip-0.8\wd0\hbox{$#1$}
}

\theoremstyle{plain}

\newtheorem{theorem}{Theorem}

\newtheorem{definition}{Definition}

\newtheorem*{claim*}{Claim}

\newtheoremstyle{named}{}{}{\itshape}{}{\bfseries}{.}{.5em}{#1\thmnote{
    #3}}
\theoremstyle{named}

\newtheoremstyle{named2}{}{}{\itshape}{}{\bfseries}{:}{.5em}{#1\thmnote{
    #3}}
\theoremstyle{named2}
\newtheorem*{namedaxiom2}{}

\theoremstyle{remark}

\begin{document}
% Page header
%\markboth{Echenique}{Revealed Preference}

\title[Risk, Uncertainty, Intertemporal choice]{New developments in revealed preference theory: decisions under
  risk, uncertainty, and intertemporal choice.}

\thanks{Division of the Humanities and Social Sciences,
  California Institute of Technology, Pasandea CA 91125; email:
  fede@caltech.edu. I thank Taisuke Imai, Kota Saito and John Quah for
  comments on an earlier draft, and the National Science Foundation
  for support through the grants SES-1558757 and CNS-1518941. This
  paper has been accepted for publication in Volume 12 of the Annual Review of
  Economics. When citing this paper, please use the following:
  Echenique FE. 2019. ``New developments in revealed preference
  theory: decisions under risk, uncertainty, and intertemporal
  choice,'' Annu.\ Rev.\ Econ.\ 12: Submitted. DOI: https://doi.org/10.1146/annurev-economics-082019-110800.}

%Authors, affiliations address.
\author[Echenique]{Federico Echenique}

%Abstract
\begin{abstract}
This survey reviews recent developments in revealed preference theory. It discusses the testable implications of theories of choice that are germane to specific economic environments. The focus is on expected utility in risky environments; subjected expected utility and maxmin expected utility in the presence of uncertainty; and exponentially discounted utility for intertemporal choice. The testable implications of these theories for data on choice from classical linear budget sets are described, and shown to follow a common thread. The theories all imply an inverse relation between prices and quantities, with different qualifications depending on the functional forms in the theory under consideration.
\end{abstract}

%Keywords, etc.
%\begin{keywords}
%revealed preference theory, expected utility, uncertainty, ambiguity, discounted utility.
%\end{keywords}
\maketitle

%Table of Contents
\tableofcontents

\maketitle

\section{Introduction}

Revealed preference theory has a long and distinguished history in economics, giving empirical meaning to the hypothesis of rationality in neoclassical economics. The investigations of \cite{samuelson1938note}, \cite{houthakker1950revealed}, \cite{afria67}, \cite{diewert1973afriat}, and \cite{varian1982nonparametric} clarified the precise behavioral meaning of the hypothesis that a consumer maximizes utility. The theory culminates in the celebrated Afriat's theorem (Theorem~\ref{thm:Afriat} below), which states that the empirical content of rational behavior (utility maximization by a locally non-satiated consumer) is the same as that of a consumer with a monotone increasing and concave utility function. In turn, the datasets that are consistent with the theory are those that satisfy Afriat's ``Generalized Axiom of Revealed Preference'' (GARP; the terminology is due to \cite{varian1982nonparametric}).\footnote{There are a number of existing surveys of the revealed preference literature: see \cite{varian2006revealed,Carvajal04b,cherchye-revealed,crawforderockempiricalRP, chambers2016revealed}.} 

While the older literature deals with general utility maximization, recent developments in revealed preference theory have sought to
understand the empirical content of specific theories of rational choice, in specific economic environments. This work is the focus of
the present survey. I focus on decisions under risk, uncertainty, and intertemporal choice. Within each of these domains,
there are theories that stand out as ubiquitous in economic modeling: expected utility theory for choice under risk, subjective expected
utility for choice under uncertainty, and exponential discounting for intertemporal choice.  

I shall describe how the GARP needs to be strengthened to capture each of the specific theories in question. For most of the discussion, I will consider the theories under the additional assumption that preferences are convex. As a consequence, first-order conditions suffice to characterize optimizing behavior, which affords some significant simplifications. 

We shall see that the relevant strengthenings of GARP all have the flavor of a negative relation between prices and quantities. For example, a risk-averse expected-utility agent who chooses to consume more conditional on state $s_1$ than conditional on state $s_2$, must either think that $s_1$ is more likely than $s_2$, or face cheaper prices for consumption conditional on state $s_1$ than on $s_2$. Such simple statements can be combined to provide a precise characterizations of risk-averse expected utility theory. More generally, all the theories under consideration imply that the data display a negative relation between prices and quantities under certain qualifications that can be traced to the functional form assumed by the theory. 

The theories under consideration are heavily used in economics. The workhorse model in macroeconomics and finance, for example, assumes expected utility theory for environments under risk and uncertainty, and discounted utility for intertemporal models. The maxmin model has more recently been used in macroeconomics \citep{hansen2008robustness}. It is also worth mentioning that the revealed preference theory of utility maximization that emerges from Afriat's theory has been extended in various ways, and applied empirically in many different settings. See for example \cite{matzkin1991testing}, \cite{matzkin1994restrictions} and \cite{blundell2003nonparametric}; or more generally \cite{chambers2016revealed} for an extensive review of these extensions and applications.

The survey is structured as follows. Section~\ref{sec:model} introduces the model and basic definitions and notational
conventions. In~\ref{sec:afriat} I present as background a discussion of general utility maximization and  Afriat's theorem. Then 
Section~\ref{sec:addsep} treats the case of additively-separable utility as warm-up, and because many of  the remaining theories will
follow similar ideas. Then in~\ref{sec:riskonegood} I turn to decisions under risk, and expected utility
theory. Section~\ref{sec:uncertaintyonegood} deals with uncertainty and subjective expected utility, while Section~\ref{sec:maxmin}
considers the maxmin model. Then~\ref{sec:polissonquah} presents a result on expected utility without assuming risk
aversion. Section~\ref{sec:PS} provides a necessary condition for data to be consistent with probabilistic sophistication. I turn to
intertemporal choice in Section~\ref{sec:intertemporalchoice}, and consider the extensions of expected utility and discounted utility to
multiple goods per period or state in Section~\ref{sec:multiplegoods}. Finally, Section~\ref{sec:conclusion} concludes.

\section{The model}\label{sec:model}

\subsection{Notational conventions and basic definitions.}

For vectors $x,y\in\Re^n$, $x\leq y$ means that $x_i\leq y_i$ for all $i=1,\dots, n$; $x < y$ means that $x\leq y$ and $x\neq y$; and $x\ll y$ means that $x_i < y_i$ for all $i=1,\dots, n$. The set of all $x\in \Re^n$ with $0\leq x$ is denoted by $\Re^n_+$, and the set of all $x\in \Re^n$ with $0\ll x$ is denoted by $\Re^n_{++}$. 

Next, consider given a finite sequence $\{x^k:k=1,\ldots,K\}$ in $\Re^n$. For each $k$ we have a collection of real numbers $x^k_1,\ldots, x^k_n$, and we can imagine forming pairs of elements drawn from \[ 
\{x^k_l: 1\leq k\leq K; 1\leq l\leq n \}.\] Two properties of such sequences of pairs will be important, and their importance will become clear later on. 

\begin{definition}
A sequence of pairs $(x^{k_i}_{l_i},x^{k'_i}_{l'_i})_{i=1}^I$ is \df{balanced} if each $k$ appears as $k_i$ (on the left of the pair) the same  number of times it appears as $k'_i$ (on the right). 
\end{definition}

\begin{definition}
A balanced sequence of pairs $(x^{k_i}_{l_i}, x^{k'_i}_{l'_i})_{i=1}^I$ is \df{doubly balanced} if each $l$  appears as $l_i$ (on the left of the pair) the same  number of times it appears as $l'_i$ (on the right). 
\end{definition}

\subsection{The model}

I shall exclusively consider the neoclassical theory of the consumer. The theory specifies as primitive a consumption space $X$, which is a subset of some Euclidean space, and a collection $\U$ of utility functions $u:X\rightarrow \Re$. Throughout this survey, $X$ will be $\Re^n_+$, the positive orthant of $\Re^n$. I shall deal with risk, uncertainty, and intertemporal tradeoffs simply by adopting different interpretations for the vectors in $\Re^n_+$. In the abstract, the vector $x=(x_1,\ldots,x_n)\in\Re^n_+$ involves consumption of $x_l$ units of ``good'' $l$. By interpreting the word ``good'' we can capture risk, uncertainty, and intertemporal choice.\footnote{\cite{debreu1959theory} contains a well-known discussion of such interpretations.}

In each case we shall appropriately  restrict $U$ to capture the most commonly considered theories of choice under risk and uncertainty, or intertemporal choice. To sum up, then, a pair $(X,U)$ describes the theory under consideration; and for the purpose of this survey I shall take $X$ to equal the positive orthant of some Euclidean space.  

The second order of business is to specify what is assumed observable: what is the available data. I shall focus on choice data drawn from the  neoclassical theory of the consumer. Specifically:

\begin{definition} A \df{dataset} is a finite collection of pairs $(x,p)\in
\Re^n_{+}\times \Re^n_{++}$. \end{definition}

The source of a dataset is a consumer, whom we have recorded making a finite collection of choices. In actual empirical applications of the theory, the data can come from consumption surveys drawn from the field, from a laboratory experiment, or from a hybrid design. 

Each choice consists of an element $x\in\Re^n_+$ selected from a \df{budget} \[ 
B(p,I) = \{z\in\Re^n_+ : p\cdot z\leq I \};
\] where $p$ is a \df{price vector}, and $I=p\cdot x$ is \df{income}. Note that $(x,p)$ describes all we need to know
about the choice and about the budget. The vector $p$ is the relevant price
vector, and $p\cdot x$ is the relevant income. Of course, this
presumes that expenditure exhausts income: some times this assumption can be questionable.

Next, we spell out the empirical content of a theory. The empirical
content consists of all the datasets that are consistent with the
theory.\footnote{See \cite{chambers2014axiomatic} for a formal study
  of empirical content.} Given a collection of utility functions $\U$,
we say that a data set $(x^k,p^k)_{k=1}^K$  is \df{$\U$-rational} if
there exists $U\in \U$ such that, for each~$k$,
\[ x^k\in\argmax\{U(x): x\in B(p^k,p^k\cdot x^k) \}.
\] 

\subsubsection{General utilities}\label{sec:afriat} For context, I give a quick overview of the most important result in revealed preference theory: Afriat's theorem.

To this end, let $\U_{\mathit{LNS}}$ be the set of locally non-satiated utility functions, and $\U_{\mathit{MC}}$ be the set of strictly monotonic and concave utility functions. 

Given a dataset $(x^k,p^k)_{k=1}^K$, we can define two binary relations on $\Re^n_+$. First, say that $x\RPref y$, ``$x$ is revealed preferred to $y$,'' if there exists $k$ such that $x=x^k$, and $p^k\cdot y\leq p^k\cdot x^k.$ Second, say that that $x\RPrefS y$, ``$x$ is strictly revealed preferred to $y$,''  if there exists $k$ such that $x=x^k$, and $p^k\cdot y< p^k\cdot x^k.$ 

A dataset $(x^k,p^k)_{k=1}^K$ satisfies the \df{Weak Axiom of Revealed Preference (WARP)} if there  is no pair of observations $k$ and $k'$ such that $x^k\RPref x^{k'}$ while $x^{k'}\RPrefS x^k$. A dataset that violates WARP cannot be $\U_{\mathit{LNS}}$ rational. 

A dataset $(x^k,p^k)_{k=1}^K$ satisfies the \df{Generalized Axiom of
  Revealed Preference (GARP)} if, for any finite sequence
$(k_i)_{i=1}^M$ in  $\{1,\ldots,K\}$, if $x^{k_i}\RPref x^{k_{i+1}}$
for $i=1,\ldots,M-1$,  then it is false that $x^{k_M}\RPrefS x^{k_1}$. 

\begin{theorem}\label{thm:Afriat}
Let $(x^k,p^k)_{k=1}^K$ be a dataset. The following statements are equivalent.
\begin{enumerate}
    \item The dataset is $\U_{\mathit{LNS}}$-rational.
    \item The dataset satisfies GARP.
    \item\label{it:AIAthm} There are scalars $V^k$ and $\la^k>0$, for $1\leq k\leq K$, such that \[V^k\leq V^{k'} + \la^{k'} p^{k'}\cdot (x^k - x^{k'}) \]  for $1\leq k,k'\leq K$.
    \item The dataset is $\U_{\mathit{MC}}$-rational.
\end{enumerate}
\end{theorem}

Theorem~\ref{thm:Afriat} is due to \cite{afria67} (see also \cite{diewert1973afriat} and \cite{varian1982nonparametric}). A proof (in fact, two proofs) and a detailed discussion of Afriat's theorem can be found in \cite{chambers2016revealed}. The system of linear inequalities in Statement~\eqref{it:AIAthm} is termed ``Afriat inequalities,'' and can be seen as fitting the data to the first-order conditions that characterize an optimum. In the rest of the survey, the idea of fitting the data to the relevant first-order conditions will be the source of many results.

The best-known implication of Theorem~\ref{thm:Afriat} is that if a dataset is $\U_{\mathit{LNS}}$-rational, then it is also $\U_{\mathit{MC}}$ rational. In other words, the hypothesis that a consumer chooses as if maximizing a concave utility has the same empirical content as the seemingly much laxer hypothesis of maximizing a locally non-satiated utility.

\subsubsection{Specific utilities}
In this survey, I shall focus on choice under risk and uncertainty, and intertemporal choice. For the case of risk and uncertainty, I suppose that there is a finite set $S$ of \df{states of the world}; let $\abs{S}=n$.  In this case, a consumption vector $x\in\Re^n_+$ consists of a state-contingent payoff. There is a single physical good; think of this good as money, and the state-contingent payoff is in units of money. Put differently, the consumer chooses a delivery $x\in\Re^S_+$ of money that depends on the state. The consumer could have access to  financial markets in which a full set of Arrow-Debreu securities are traded, and obtain $x$ as the payoff derived from a portfolio of Arrow-Debreu securities. The price vector $p$ reflects the prices of the securities, and income $I=p\cdot x$ is how much the consumer can afford to spend. 

For intertemporal choice, I shall assume that there are $T$ time periods ($T=n$), and $x\in\Re^T_+$ represents  consumption over time. Again, there is a single good, termed ``money,'' and $x\in\Re^T_+$ is a \df{consumption path}, representing dated quantities of consumption. 

\section{Additive separability}\label{sec:addsep}

As an introduction to the main ideas in the survey, I shall consider the theory of additively separable utility. In the case of intertemporal choice, additive separability coincides with no-discounting stationary utility (as analyzed, for example, by \cite{browning1989nonparametric}, but more on that later in the survey). In the case of choice under risk or  uncertainty, a consumer with additive utility has a uniform prior over states. All states are equally likely.

Let $\U_{\mathit{AS}}$ be the class of utility functions $U:\Re^n_+\rightarrow \Re$ for which there exists a concave and strictly increasing
$u:\Re_+\rightarrow \Re$ such that  $U(x)\geq U(y)$ if and only if  \[ 
\sum_{l=1}^n  u(x_l)  \geq \sum_{l=1}^n u(y_l).
\] 

Consider the problem that a $\U_{\mathit{AS}}$-rational consumer is supposed to be solving, or acting {\em as if} she is solving:
\[ 
\max \{\sum_{l=1}^n u(x_l) : x\in B(p,I)\}.
\] Suppose, for the purpose of exposition, that $u$ is smooth. Then the first-order conditions of the maximization problem (assuming an interior solution) imply that \[ 
\frac{u'(x_l)}{u'(x_{l'})} =\frac{p_l}{p_{l'}}.
\] The first-order condition, and the  concavity of $u$, imply that whenever $x_l > x_{l'}$ it must be the case that $\frac{p_l}{p_{l'}}\leq 1.$ In words, larger consumption of good $l$ than of $l'$ is only possible when $l$ is cheaper than $l'$: demand ``slopes down.''

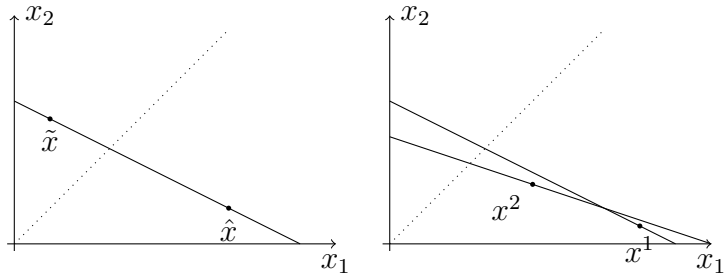
\begin{figure}
\centering
\begin{subfigure}
  \centering
  \begin{tikzpicture}[scale=.95]
\draw[->] (0,-.1) -- (0,3.2) node[anchor=west] {$x_{2}$};
\draw[->] (-.1,0) -- (4.5,0) node[anchor=north] {$x_{1}$};
\draw[dotted] (0,0) -- (3,3);

\draw[-] (0,2) -- (4,0); % p_1= 1/4, p_2=1/2  I=1 p_1/p_2=1/2
\fill (3, 2- 3/2 ) circle (1pt);% {$x$};
\path (3, 2- 3/2 ) node[anchor=north] {$\hat x$};
\fill (.5, 2- .5/2 ) circle (1pt);% {$x$};
\path (.5, 2- .5/2 ) node[anchor=north] {$\tilde x$};
\end{tikzpicture}
\end{subfigure}
\begin{subfigure}
  \centering
\begin{tikzpicture}[scale=.95]
\draw[->] (0,-.1) -- (0,3.2) node[anchor=west] {$x_{2}$};
\draw[->] (-.1,0) -- (4.5,0) node[anchor=north] {$x_{1}$};
\draw[dotted] (0,0) -- (3,3);

\draw[-] (0,2) -- (4,0); % p_1= 1/4, p_2=1/2  I=1 p_1/p_2=1/2
\fill (3.5, 2- 3.5/2 ) circle (1pt);% {$x$};

\draw[-] (0,3/2) -- (9/2,0); % p'_1= 1/9, p'_2=1/3 I=1/2 p_1/p_2=1/3
\fill (2, 3/2- 2/3 ) circle (1pt);% {$x$};

\path (3.5, 2- 3.5/2 ) node[anchor=north] {$x^{1}$};
\path (2, 3/2- 2/3 ) node[anchor=north east] {$x^{2}$};
\end{tikzpicture}
\end{subfigure}
\caption{Additively separable utility}
\label{fig:addsep}
\end{figure}

As an example, consider the diagram on the left in Figure~\ref{fig:addsep}. It depicts a budget set where good one is cheaper than good two: the budget set below the 45-degree line is larger than that above. If a dataset would include a choice of $\tilde x$ on this budget set, it would violate downward-sloping demand and not be $\U_{\mathit{AS}}$-rational. Choices such as $\hat x$ are, in and of themselves, compatible with $\U_{\mathit AS}$. But we shall see that, together with other observations in a dataset, they can still trigger a rejection of the theory of additively separable utility.

So the theory of additively separable utility has implications for the data, a property that we think of as  downward-sloping demand: it says that larger price are associated with smaller quantities.\footnote{I use the term ``downward-sloping demand'' for lack of a better one. The prices and quantities that are being compared correspond to different goods (or states, or time-periods in later interpretations). In contrast, the term is most often used to describe a comparative statics property of demand, not how prices and quantities compare at a given price vector.} We shall express this downward-sloping demand property in a way that turns out to be useful for the rest of our analysis:

\begin{definition}
A sequence of pairs $(x^{k_i}_{l_i}, x^{k'_i}_{l'_i})_{i=1}^I$ has the \df{downward-sloping demand property}  if \[ x^{k_i}_{l_i} > x^{k'_i}_{l'_i} \text{ for all } i \text{ implies that } \prod_{i=1}^n \frac{p^{k_i}_{l_i}}{p^{k'_i}_{l'_i}} \leq 1.
\]
\end{definition} 

The precise property of datasets we are interested in can be called the

\begin{namedaxiom2}[Strong Axiom of Revealed Additively Separable Utility]
Any balanced sequence of pairs has the risk-neutral  downward-sloping property. 
\end{namedaxiom2}

The Strong Axiom of Revealed Additively Separable Utility is a test for whether a dataset is $\U_{\mathit{AS}}$-rational.

\begin{theorem}\label{thm:additivesep}
  A dataset is $\U_{\mathit{AS}}$ rational if and only if it satisfies the Strong Axiom of Revealed Additively Separable Utility
\end{theorem}

I shall not prove Theorem~\ref{thm:additivesep}, or any other results stated in this survey, but a proof can be obtained easily using the technique applied to prove Theorem~\ref{thm:thetheorem3} below.\footnote{See the references cited for Theorem~\ref{thm:thetheorem3}.} I should also note that more general notions of separability is treated in \cite{varian1983non} and \cite{quahseparability}.

The role of a balanced sequence of pairs in Theorem~\ref{thm:additivesep} needs to be explained. Consider the diagram on the right of Figure~\ref{fig:addsep}. It depicts a dataset with  two observations, $(p^1,x^1)$ and $(p^2,x^2)$, that violate the weak axiom of revealed preference (WARP). Hence it cannot be rationalized by any utility function, let alone by a member of $\U_{\mathit{AS}}$. Theorem~\ref{thm:additivesep} implies that this dataset must exhibit some kind of violation of downward-sloping demand. 

If we had a single observation: $K=1$, then our discussion of downward-sloping demand is straightforward: $\frac{u'(x_l)}{u'(x_{l'})} =\frac{p_l}{p_{l'}}$ and concavity implies that quantities and prices must move in opposite directions. In each of  the two observations on the right of Figure~\ref{fig:addsep} we have that good one is cheaper than good two, and sure enough the chosen consumption bundle involves larger quantities of good one. Thus each observation {\em alone} is $\U_{\mathit{AS}}$-rational.

However, when there is more than one observation in a dataset, it is possible to ``mix and match'' pairs from different observations. In particular, in Figure~\ref{fig:addsep} we have \[ x^1_1>x^2_{1} \text{ and  } x^2_{2}>x^1_{2} \text{ while } \frac{p^1_1}{p^2_{1} }\frac{p^2_2}{x^1_{2} }>1. \] In this sense, the data violates the downward-sloping demand property. The sequence of pairs $(x^1_1,x^2_1),(x^2_2,x^1_2)$ is balanced, and it contradicts the downward-sloping demand condition.

The source of this imposition on the data is again the first-order
condition $\frac{u'(x_l)}{u'(x_{l'})} =\frac{p_l}{p_{l'}}$. If the
data were $\U_{\mathit{AS}}$-rational we would need a utility function
that satisfies  this first-order condition. Then we would have to
have \[1\geq \frac{u'(x^1_1)}{u'(x^2_{1})}
  \frac{u'(x^2_2)}{u'(x^1_{2})}   =\frac{p^1_l}{p^2_{1}}
  \frac{p^2_2}{p^1_{2}},\] which is violated by the data in Figure~\ref{fig:addsep}.

``Mix and match'' is where the balancedness of sequences comes in. For the theory to have definitive implications on the relation between quantity and price, we need to include $k$ on the left of a pair exactly as many times as we include it on the right.\footnote{The axioms in \cite{fudenberg2013stochastic} present similar issues. See also the treatment of additively separable utility in \cite{fishburn1970utility}.} 

\section{Choice under risk and uncertainty}

To study choice under risk and uncertainty we think of consumption vectors $x$ as representing state-contingent monetary payoffs. Consider a finite set $S$ of \df{states of the world} with $\abs{S}=n$, and interpret $x=(x_1,\ldots, x_n)\in\Re^n_+$ as a monetary payment that depends on the state. The vector $x$ pays $x_i$ if the $i$'th state occurs.

Let $\Delta_{++} = \{\mu \in \Re^n_{++} : \sum_{s=1}^n \mu_s =1\}$
denote the set of strictly positive probability measures on $S$.
When we talk about risk, we assume that there is given a known
(``objective'') probability measure over states. Risk is common in
experimental designs, such as those of \cite{choi2007},
\cite{choi2014more}, and \cite{halevy2018parametric}. \footnote{The
  experimental literature on risk and uncertainty is extensive. See
  the recent survey by  \cite{trautmann2015ambiguity}. I restrict
  attention to experiments that produce the type of datasets assumed
  in this survey. That is, experimental designs that focus on choices
from budgets.}

In situations of uncertainty, in contrast, there is no given probability measure. Instead, we shall let the probabilities form part of the preferences that should be explaining the data. Uncertainty is of course prevalent in data from the field. Many lab experiments induce an objective probability, for example \cite{choi2007} and \cite{choi2014more}. There are, however,  experimental designs with uncertainty, such as \cite{heypace} and \cite{ourseuexp}.

\subsection{Risk: Expected utility}\label{sec:riskonegood}

Suppose that a probability measure $\mu^*\in \Delta_{++}$ is given and known. Think of $\mu^*$ as an ``objective'' probability; one resulting from a known and objective randomization device. Such devices are regularly used in laboratory experiments, and the result that I am about to present is applicable to data drawn from such laboratory experiments.\footnote{See \cite{approximateEU} for an application of this theory to experimental data.}

Let $\U_{\mathit{EU}}$ be the class of utility functions $U:\Re^n_+\rightarrow \Re$ for which there exists  a concave and strictly increasing $u:\Re_+\rightarrow \Re$ such that 
$U(x)\geq U(y)$ if and only if  \[
\sum_{s\in S}  \mu^*_s u(x_s)  \geq \sum_{s\in S} \mu^*_s u(y_s).
\] 

In studies of objective expected utility, the concept of price-probability ratios, or
``risk-neutral prices,'' is crucial. Risk-neutral prices are defined
as $\rho^k_s= p^k_s/\mu^*_s$, for $k \in K$ and $s \in S$. To understand the role of risk-neutral prices, note that the first-order conditions for the consumer's optimization problem imply that $u'(x_s)/u'(x_{s'}) = \rho_s/\rho_{s'}$. Compare this equation with the first-order condition for additive separability in our previous discussion: after transforming prices into risk-neutral prices, the analysis can proceed exactly as in Section~\ref{sec:addsep}.

In consequence, the relevant property is

\begin{namedaxiom2}[Strong Axiom of Revealed Objective Expected Utility (SAROEU)] Any balanced sequence of pairs has the risk-neutral  downward-sloping property.  \end{namedaxiom2}

We obtain a characterization of $\U_{\mathit{EU}}$-rational datasets. 

\begin{theorem}\label{thm:thetheorem3} A dataset is $\U_{\mathit{EU}}$ rational if and only if it satisfies SAROEU.
\end{theorem}

Theorem~\ref{thm:thetheorem3} is due to \cite{kubler2013}. The appendix to \cite{seu} considers a hybrid case, where some states have subjective and known probabilities, while others do not. The hybrid case is important to cover data from experiments like those of \cite{choi} and \cite{bossaerts2010ambiguity}.

\subsection{Uncertainty: Subjective Expected Utility.}\label{sec:uncertaintyonegood} 

We shall be interested in the theory of subjective expected utility,
the most commonly used theory of choice under uncertainty. 

Let $\U_{\mathit{SEU}}$ be the class of utility functions $U:\Re^n_+\rightarrow \Re$ for which there exists  $\mu\in \Delta_{++}$ and a concave and strictly increasing $u:\Re_+\rightarrow \Re$ such that $U(x)\geq U(y)$ if and only if  \[ 
\sum_{s\in S} \mu_s u(x_s)  \geq \sum_{s\in S} \mu_s u(y_s).
\] Here $\mu$ is interpreted as a subjective belief over states, while $u$ is a concave utility function over money. The assumption that $u$ is concave means that I restrict attention to risk-averse consumers.\footnote{For an analysis that avoids the concavity assumption, see Section~\ref{sec:polissonquah}.}

With subjective expected utility, the relevant first-order condition is that
\[
\frac{\mu_s u'(x_s)}{\mu_{s'}u'(x_{s'})} = \frac{p_s}{p_{s'}}.\] A dataset like the one with $\tilde x$ on the left of Figure~\ref{fig:addsep} is $\U_{\mathit{SEU}}$-rational because an agent might choose to consume more in the more expensive state because she thinks it is the more likely state. We do not have access to probabilities, which leaves us with fewer constraints than in $\U_{\mathit{EU}}$: probabilities may be adjusted in constructing a rationalizing member of $\U_{\mathit{SEU}}$.

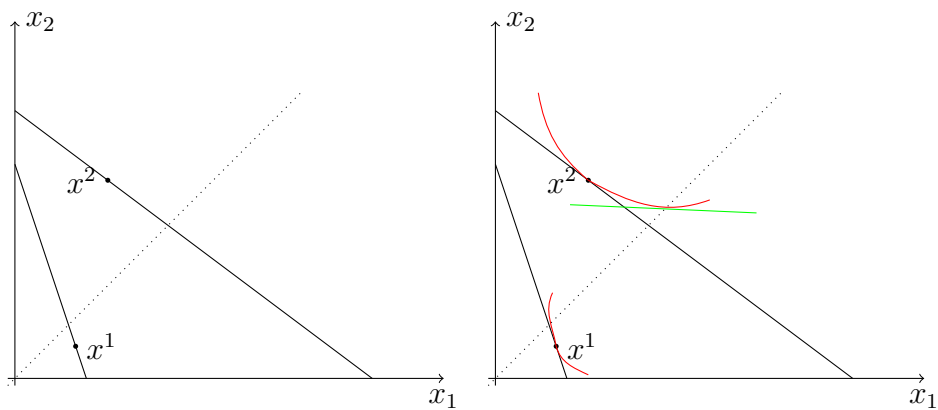
\begin{figure}
\centering
\begin{subfigure}%{.4\textwidth}
  \centering
\begin{tikzpicture}[scale=.95]
\draw[->] (0,-.1) -- (0,5) node[anchor=west] {$x_{2}$};;
\draw[->] (-.1,0) -- (6,0) node[anchor=north] {$x_{1}$};;
\draw[dotted,thin] (-.1,-.1) -- (4,4);

\draw[-] (0,3) -- (1,0);          
\fill (.85, 3- .85*3)  circle (1pt);

\draw[-] (0,15/4) -- (15/3,0);          
\fill (1.3, 15/4 - 3.9/4) circle (1pt);

\path (.85, 3- .85*3) node[anchor=west] {$x^{1}$}
(1.3, 15/4 - 3.9/4) node[anchor=east] {$x^{2}$};
\end{tikzpicture}
\end{subfigure}
\begin{subfigure}%{.4\textwidth}
  \centering
\begin{tikzpicture}[scale=.95]
\draw[->] (0,-.1) -- (0,5) node[anchor=west] {$x_{2}$};
\draw[->] (-.1,0) -- (6,0) node[anchor=north] {$x_{1}$};
\draw[dotted,thin] (-.1,-.1) -- (4,4);

\draw[-] (0,3) -- (1,0);          
\fill (.85, 3- .85*3)  circle (1pt);

\draw[-] (0,15/4) -- (15/3,0);          
\fill (1.3, 15/4 - 3.9/4) circle (1pt);

\path (.85, 3- .85*3) node[anchor=west] {$x^{1}$}
(1.3, 15/4 - 3.9/4) node[anchor=east] {$x^{2}$};

\draw[-,red] (.8,1.2) to [out=-115,in=100]  (.85, 3- .85*3);
\draw [-,red] (.85, 3- .85*3) to [out=-75,in=155] (1.3,.05);

\draw[-,red] (.6,4) to [out=-80,in=140]  (1.3, 15/4 - 3.9/4);
\draw [-,red] (1.3, 15/4 - 3.9/4) to [out=-29,in=200] (3,2.5);

\draw[-,green] (1.044*1, 1.044*2.33) -- (1.044*3.5, 1.044*2.22 );
\end{tikzpicture}
\end{subfigure}
\caption{Violation of subjective expected utility}
\label{fig:seu}
\end{figure}

So what are the $\U_{\mathit{SEU}}$-rational data? Consider the data on the left of Figure~\ref{fig:seu}. It is easy to see that the data is not $\U_{AS}$-rational, but that does not help us very much because $\U_{\mathit{AS}}$ is a more restrictive theory than $\U_{\mathit{SEU}}$. It is also easy to see that the data satisfies the weak axiom of revealed preference, and is therefore rational for {\em some} utility function (it is $\U_{\mathit{LNS}}$-rational). Of course, this does not tell us whether the rationalizing utility could be a member of $\U_{\mathit{SEU}}$. The way to make progress turns out to be to consider the 45-degree line.

Suppose that the dataset in Figure~\ref{fig:seu} were $\U_{\mathit{SEU}}$-rational. The rationalizing utility would then satisfy the tangency properties at $x^1$ and $x^2$ depicted in the diagram on the right of Figure~\ref{fig:seu}. The tangencies express the first-order conditions I have been talking about. Now consider the indifference curve at $x^2$, and follow it until it crosses the 45-degree line. At the crossing point, the tangent line to the indifference curve must be flatter than the tangent line at $x^2$: a line that coincides with the budget line at $x^2$. The reason for why it is flatter is that $x^2$ lies above the 45-degree line. On the other hand, and symmetrically, the indifference curve at $x^1$ has a tangent at the 45-degree line that is steeper than the budget line at $x^1$. 

Since the prices at $x^1$ imply a steeper budget line than at $x^2$, we are forced to conclude that the tangent,  on the 45-degree line,  to the indifference curve passing through $x^1$ is {\em steeper} than the same tangent for the indifference curve passing through $x^2$. So the existence of a rationalizing member of $\U_{\mathit{SEU}}$ implies that the tangent to the lower indifference curve at the 45-degree line is steeper than that at the higher indifference curve.

That is, however, impossible if the data is $\U_{\mathit{SEU}}$-rational because anywhere on the 45-degree line we have $x_1=x_2$, so the tangent has slope \[\frac{\mu_1 \cancel{u'(x_1)}}{\mu_{2}\cancel{u'(x_{2})}} = \frac{\mu_1}{\mu_{2}}.\] The slope of the tangent is the same anywhere on the 45-degree line, which contradicts what we just concluded from the data on Figure~\ref{fig:seu}.

The relevant property of the data turns out to be the following axiom.

\begin{namedaxiom2}[Strong Axiom of Revealed Subjective Expected Utility] Any doubly balanced sequence of pairs has the downward-sloping demand property.  \end{namedaxiom2}

\begin{theorem}\label{thm:thetheorem2}
A dataset is $\U_{\mathit{SEU}}$ rational if and only if it satisfies the Strong Axiom of Revealed Subjective Expected Utility.
\end{theorem}

Theorem~\ref{thm:thetheorem2} is due to \cite{seu}.

\subsection{Uncertainty: Maxmin Expected Utility}\label{sec:maxmin}

Subjective expected utility struggles to account for the phenomenon of aversion to ambiguity: people seem to make choices that are inconsistent with fixed and stable subjective beliefs. Their choices seem to reflect an aversion to equating an uncertain, ambiguous, situation to one with given probabilistic beliefs.\footnote{The evidence usually offered in support of ambiguity aversion is the Ellsberg paradox, sue to \cite{ellsberg1961}. See \cite{camerer1992recent} for a survey.} The leading theory of choice that accounts for such ambiguity aversion is (arguably) the theory of maxmin expected utility \citep{gilboa1989maxmin}, to which I now turn.

Let $\U_{\mathit{MEU}}$ be the class of utility functions $U:\Re^n_+\rightarrow \Re$ for which there exists   $P\subseteq \Delta_{++}$, nonempty, closed and convex, and a concave and strictly increasing $u:\Re_+\rightarrow \Re$ such that  $U(x)\geq U(y)$ if and only if  \[ \min\{ \sum_{s\in S} \mu_s u(x_s) :\mu\in P \}\geq  \min\{ \sum_{s\in S} \mu_s u(y_s) :\mu\in P \}. \]

The results on $\U_{\mathit{MEU}}$ rationality require the assumption that $\abs{S}=2$: the set of states if two. Obviously this is a very big
limitation, but the theory is still useful because many lab experiments operate with two states. So, while it would be very good
to have more general results, the ones we have remain applicable to many relevant datasets.\footnote{Including \cite{heypace}, see
  \cite{chambersPNAS}, and \cite{ourseuexp}. In fact, in the
  experiment of \cite{ourseuexp} there are three states, but the
  authors find a way to leverage the two-state result. }

We shall need some additional notation. Assuming that there are only two states, let $S=\{1,2\}$. Given is a dataset $\{(x^k,p^k) : k=1,\ldots,K\}$. Let $K_0$ be the set of all $k$ such that $x^k_1 = x^k_2$. Let $K_1$ be the set of all $k$ such that $x^k_1 < x^k_2$, and  $K_2$ be the set of all $k$ such that $x^k_1 > x^k_2$. Note that  $[K]=K_0\cup K_1\cup K_2$.

Given a sequence of pairs $(x^{k_i}_{s_i}, x^{k'_i}_{s'_i})_{i=1}^I$, consider the following notation: Let $I_{l,s} = \{i: k_i\in K_l \text{ and } s_i=s\}$, $I'_{l,s} = \{i: k'_i\in K_l \text{ and } s'_i=s\}$, for $l=0, 1,2$ and $s=1,2$. Thus equipped, the property of interest is

\begin{namedaxiom2}[Strong Axiom of Revealed Maxmin Expected Utility]\label{theaxiommaxmin} 
Any balanced sequence of pairs $(x^{k_i}_{s_i},
x^{k'_i}_{s'_i})_{i=1}^I$ in which
$ \abs{ I_{0,1} }+ \abs{ I_{1,1} } - \abs{ I'_{1,1} }  = \abs{
  I'_{0,1}}+\abs{ I'_{2,1} } - \abs{ I_{2,1} } \leq 0$ has
  the downward-sloping demand property. 
\end{namedaxiom2} 

\begin{theorem}\label{thm:maxmin}
  A dataset is $\U_{\mathit{MEU}}$ rational if and only if it satisfies the Strong Axiom of Revealed Maxmin Expected Utility.
\end{theorem}

Theorem~\ref{thm:maxmin} is due to \cite{chambersPNAS}. 

The theory of maxmin expected utility presents some challenges that were not present in expected utility theory. To rationalize a dataset we have the freedom of choosing different beliefs for different observations. Such freedom makes it difficult, in general, to obtain a tight axiomatization, but the case of two states has one significant advantage. When there are only two states, there are only two extreme points to the set of beliefs $P$. Moreover, depending on whether $x^k_1\leq x^k_2$, or $x^k_2\leq x^k_1$, we know which of the two extremes is relevant in explaining the choice of $x^k$. If it is $x^k_1\leq x^k_2$, then the relevant belief must be pessimistic about state 1. If instead $x^k_2\leq x^k_1$ then he relevant belief must be pessimistic about state 2. Theorem~\ref{thm:maxmin} is made possible by this property of the two-state case.

\cite{chambersPNAS}  present a characterization of the datasets that are rationalizable by other theories of choice, including Choquet expected utility, and MEU with $\abs{S}>2$. These results are, however, limited to risk neutrality.\footnote{The methods of \cite{polissonquahrenou}, briefly outlined below, can also be used for general theories of choice under uncertainty.}

\subsection{Expected utility without risk aversion}\label{sec:polissonquah}

The discussion so far has been restricted to models with risk aversion. In consequence, the tests I have presented are really joint tests of the hypotheses that an agent is consistent with some particular theory of choice, {\em and} that the agent is risk averse. One can, however, imagine an expected utility-rational agent, for example, failing such a test  because they make  choices that indicate a pure preference for taking on risks. These choices could for example be on the corners of the budget sets that they face.\footnote{It is easy to generate examples of datasets that are not $\U_{\mathit{EU}}$ rational as defined here, but that can be rationalized with a non-concave utility. Expected utility with risk aversion implies a normal demand function, so an example can be generated using the conditions in \cite{CHERCHYE2018361}.} I now proceed to outline a method for testing EU (and other theories of choice under risk and uncertainty) that avoids the assumption of risk aversion. The method is due to \cite{polissonquahrenou}. 

Given a dataset $(x^k,p^k)$, $1\leq k\leq K$, let $\X^* =\{x^k_s :1\leq k\leq K, 1\leq s\leq n\}$ be the set of all observed chosen consumption quantitites. Define $\X =  \{0\}\cup\X^*$, and let $\X^n$ denote the $n$-fold cartesian product $\X\times \cdots \times \X$. 

A data-set $(x^k,p^k)_{k=1}^K$ is \df{$\U_{\mathit{EU}}$-rational on the data} if there exists a strictly increasing utility function $u:\X \rightarrow \Re$ such that, for each $k$, \[ x^k\in\argmax\{U(x): x\in B(p^k,p^k\cdot x^k) \cap \X^n\}. \]

Testing whether a dataset is $\U_{\mathit{EU}}$-rational on the data amounts to checking whether a utility function can be constructed on finitely many points to satisfy a system of linear inequalities. It is therefore computationally easy to do (there are at most $(\abs{S}K)^2$ linear inequalities to check for satisfaction). 

The next result is due to \cite{polissonquahrenou}.

\begin{theorem}\label{thm:polissonquah} A dataset is $\U_{\mathit{EU}}$-rational if and only if it is
  $\U_{\mathit{EU}}$-rational on the data. 
\end{theorem}

The importance of Theorem~\ref{thm:polissonquah} is that budget sets are infinite, and therefore there is in principle no computationally feasible way of checking $\U_{\mathit{EU}}$-rationality. The result of \cite{polissonquahrenou} says that such a computational method is readily available, and amounts to checking whether a utility can be constructed on the data.\footnote{A very general result in \cite{GRPT} also implies that a computable test exists, but not that it is efficiently computable.} The results on risk aversion that I have discussed earlier in the survey avoid the issue by using the sufficiency of first-order conditions: there are finitely many of those.  

The statement in Theorem~\ref{thm:polissonquah} resembles part of Afriat's theorem, which can also be understood as saying that a dataset is rational if and only if it can be rationalized on the data.\footnote{See also \cite{quahseparability}, who presents a result that is similar in spirit, for separability in demand, and \cite{polissonsep} who studies additive separability without imposing concavity.}

Polisson et.\ al.\ provide a more general set of tools than evidenced by Theorem~\ref{thm:polissonquah}. They provide a framework for testing a general class of models of choice under risk (including models of non-expected utility such as rank-dependent utility), and uncertainty --- including SEU, but also models of ambiguity aversion. For some of these models, however, computational tractability is lost.

In addition to the binary pass/not-pass test I have discussed, Polisson et.\ al.\ develop an approach to measuring the distance between the theory and a dataset. Their approach is based on adapting Afriat's critical cost efficiency index to their test. 

\subsection{Probabilistic sophistication}\label{sec:PS}

Perhaps the most basic Bayesian model of decision under uncertainty is due to \cite{machina1992more}, who propose that an agent is \df{probabilistically sophisticated} if $x\in \Re^n_+$ is evaluated based on the distribution that it induces on $\Re_+$, given some prior $\mu$. That is, choices are as if there is some probability distribution $\mu$ such that  $x$ is preferred to $x'$ whenever the distribution $F_x(z)=\mu(\{s\in S:x_s\leq z\})$ is ranked above the distribution $F_{x'}(z)=\mu(\{s\in S:x'_s\leq z\})$. Probabilistic sophistication requires that the preferences over distributions respect monotonicity in first-order stochastic dominance, but imposes no additional restrictions. 

Formally, let $\U_{\mathit{PS}}$ be the set of utility functions $U:\Re^n_+\rightarrow\Re$ for which there exists $\mu\in\Delta_{++}$ and a function $V:D\rightarrow \Re$, where $D$ is the set of distributions on $\Re$ with finite support, and $V$ is monotone with respect to first-order stochastic dominance, such that $U(x)\geq U(x')$ iff $v(F_x)\geq V(F_{x'})$

\cite{epstein2000probabilities} works out a basic implication of probabilistic sophistication for datasets $(x^k,p^k)$. He shows that

\begin{theorem}\label{thm:epstein}
If a dataset $(x^k,p^k)$ $1\leq k\leq K$ is $\U_{\mathit{PS}}$-rational then there cannot exists $k,k'\in \{1,\ldots,K\}$ and $s,t\in S $ such that \begin{enumerate}
    \item $p^k_t\geq p^k_s$ and $p^{k'}_s\geq p^{k'}_t$, with at least one inequality being strict, 
    \item and $x^k_t> x^k_s$ and $x^{k'}_s > x^{k'}_t$.
\end{enumerate}
\end{theorem}

The idea behind Theorem~\ref{thm:epstein} is that if the agent's choices are guided by probabilities (if what matters about the choice of $x$ is only the probability distribution over money that it induces), then $p^k_t> p^k_s$ and $x^k_t> x^k_s$ means that the probability of state $t$ must be strictly greater than the probability of state $s$. The reason is that, when $p^k_t> p^k_s$, the agent could modify $x^k$ by choosing to consume $x^k_t$ in state $s$ and $x^k_s$ in state $t$: the modified vector would cost less than $x^k$, and therefore be affordable. For the modified vector not to first-order stochastically dominate $x^k$, the state $t$ must be more likely than $s$. Now, of course, having made that inference based on observation $(x^k,p^k)$, it should not be contradicted by observation $(x^{k'},p^{k'})$. 

Theorem~\ref{thm:epstein} provides a necessary condition for a dataset to be $\U_{\mathit{PS}}$-rational. It is an open problem to establish a necessary and sufficient condition (a ``revealed preference characterization'') for probabilistic sophistication. A natural question is whether probabilistic sophistication can be distinguished empirically from the more stringent theory of subjective expected utility; this question is answered in the affirmative by \cite{seu}, who exhibit an example of a dataset that is $\U_{\mathit{PS}}$-rational but not $\U_{\mathit{SEU}}$-rational.\footnote{In the case of risk, with objectively known probabilities, a test of monotonicity with respect to first-order stochastic dominance is developed by \cite{nishimura2017comprehensive} and implemented by \cite{polissonquahrenou}.}

\section{Intertemporal choice: exponential discounting}\label{sec:intertemporalchoice}

I shall be interested in the theory of exponentially discounted utility, the most widely applied theory of intertemporal choice. Let
$\U_{\mathit{EDU}}$ be the class of utility functions  $U:\Re^T_+\rightarrow \Re$ for which there exists $\delta \in (0,1]$ and $u:\Re_+\rightarrow \Re$ such that  $U(x)\geq U(y)$ if and only if  \[  \sum_{t\in T} \delta^t u(x_t) \geq
  \sum_{t\in T} \delta^t u(y_t) \] 

Note that, while they are meant to be applied to different environments, $\U_{\mathit{EDU}}\subseteq \U_{\mathit{SEU}}$. So the axiom that characterizes $\U_{\mathit{EDU}}$-rationality has to be more stringent than the Strong Axiom of Revealed Subjective Expected Utility.\footnote{In the intertemporal context, $\U_{\mathit{SEU}}$ is comprised of utilities with the form $\sum_t D(t)u(x_t)$, where $D(t)>0$ is a general discount function.} In fact, instead of a doubly-balanced sequence we shall count time periods on the left and on the right of each pair in the sequence.

\begin{namedaxiom2}[Strong Axiom of Revealed Exponentially Discounted  Utility] Any balanced sequence of pairs 
$(x^{k_i}_{t_i}, x^{k'_i}_{t'_i})_{i=1}^I$ for which $\sum_{i=1}^I t_i \ge \sum_{i=1}^I t'_i$, has the downward-sloping demand property.   \end{namedaxiom2}

The Strong Axiom of Revealed Exponentially Discounted  Utility has a simple explanation. Suppose, for example, that one observation in the dataset satisfies $x^k_3>x^k_2$, meaning that the consumer chooses larger consumption in period 4 than in period 3. This can only be compatible with {\em discounted} utility (which implies an impatient consumer) if consumption in period 4 is cheaper than in period 3. In contrast, $x^k_2>x^k_3$ is possible regardless of how prices compare, because the consumer's discount factor could account for  her desire to consume early, even when it is more expensive. More generally, $\sum_{i=1}^n t_i \ge \sum_{i=1}^n t'_i$ says that the consumption quantities $x^{k_i}_{t_i}$ occur later in time than the quantities $ x^{k'_i}_{t'_i}$. If, in any pair, the former are always lager than the latter, the only explanation must be a corresponding movement in prices. 

The result, obtained in  \cite{edu}, is that:

\begin{theorem} \label{thm:edu}
A dataset is $\U_{\mathit{EDU}}$ rational if and only if it satisfies Strong Axiom of Revealed Exponentially Discounted  Utility.
\end{theorem}

It is possible to obtain many results in the same spirit as Theorem~\ref{thm:edu}. \cite{edu} gives revealed preference axioms for many other theories of intertemporal choice, including quasi-hyperbolic discounting, and monotone time discounting. Echenique et.\ al.\ implement these tests on data drawn from experiments by \cite{andreonisprenger} and \cite{carvalho2014poverty}.

There is a long literature on the testable implications of models of intertemporal choice. I want to mention two papers that follow in the revealed preference tradition.  \cite{DZIEWULSKI201867} also considers exponentially discounted utility and its generalizations, but focuses on datasets that arise from pairwise comparisons; not choice from budgets. \cite{abiadams} is a study of intertemporal choice and dynamic consistency, but their analysis of household consumption is outside the (narrow) scope of my survey. 

\section{Multiple physical goods}\label{sec:multiplegoods}

One of the limitations of the results we have presented is that they are restricted to the case of a single physical good (``money'') in each time period or state. As a consequence, the results are applicable to common economic environments where agents choose dated, or state-contingent, monetary payoffs. It is, however, easy to  envision economic applications with multiple physical goods, and where the multiplicity is a crucial component of the environment.  An important application is to data from consumption surveys, where an agent or a household, records their purchases of multiple consumption goods over time. The methods I have discussed so far would require that one aggregate the multiple goods into one. Such aggregation  is complicated, both conceptually and practically.\footnote{That said, most revealed preference studies of data from consumption surveys do some sort of aggregation. Different kinds of meat or cheese are comprised into aggregate ``meat good,'' ``cheese good,'' and so on. See \cite{chambers2016revealed} for an overview of this issue.}

I shall briefly outline some revealed-preference studies that avoid the assumption of a single physical goods. 

\subsection{Risk} Suppose that there are $L$ physical goods, and $S$ states of the world. Consumption space is therefore $\Re^{LS}_+$, with $n=LS$ being the number of goods. As in Section~\ref{sec:riskonegood}, suppose given a probability measure $\mu^*\in \Delta_{++}\subseteq\Re^S_+$. Let $\U_{\mathit{EU}}$ be the class of utility functions $U:\Re^{LS}_+ \rightarrow \Re$ for which there exists a concave and strictly increasing $u:\Re^L_+\rightarrow \Re$ such that $U(x)\geq U(y)$ iff \[ 
\sum_{s\in S}  \mu^*_s u(x_s)  \geq \sum_{s\in S} \mu^*_s u(y_s).
\] 

A data set is, again, a collection $(x^k,p^k)$, $1\leq k \leq K$. It is now useful to collect all the observed quantities in a set: Let $\X=\{x_{s}^k\in \Re^L_+: 1\leq k\leq K, 1\leq s\leq S\}$. Denote by $\Delta(\X)$ the set of all probability distributions on $\X$ (a finite set). 

Recall at this point the revealed preference binary relations I introduced in Section~\ref{sec:afriat}. We shall need to consider new revealed preference relations on $\Delta(\X)$, but these build on $\RPref$ and $\RPrefS$. Say that $x\Rchris z$ if there exists $z'\in\Delta(\X)$ such that  $x\RPref z'$, and that $z$ is a mean-preferving spread of $z'$. Similarly, say that $x\RchrisS z$ if there exists $z'\in\Delta(\X)$ such that $x\RPrefS z'$, and $z$ is a mean-preserving spread of $z'$.\footnote{The idea of ``composing'' the revealed preference relation with a given (fixed) binary relation to reflect a monotonicity property is common in revealed preference theory. See, for example, \cite{chambers2009supermodularity} and \cite{nishimura2017comprehensive}.}

Next, Theorem~\ref{thm:chrisEU} is a composite of contributions by \cite{varian1983}, \cite{green1986}, and \cite{chambers2016EUtest}.\footnote{See also \cite{diewert2012afriat}  and \cite{varian1988}.}

\begin{theorem}\label{thm:chrisEU}
The following statements are equivalent:
\begin{enumerate}
    \item\label{it:Chris1} For any sequence $\{z^k\}_{k=1}^K$ in $\Delta(\X)$ with $x^k \Rchris z^k$, and any vector $\pi\in\Re^K_+$ with $\sum_k \pi^k=1$, if $\sum_k \pi^k x^k = \sum_k \pi^k z^k$, then $\pi^k=0$ for all $k$ with $x^k \RchrisS z^k$.
    \item\label{it:Chris2} For all functions $h^k_s:K\times S\rightarrow \Re_+$ with the property that \[\mu^*_s = 
    \sum_{k'}\sum_{s'} h^k_s (k',s') = \sum_{k'}\sum_{s'} h^{k'}_{s'} (k,s)\] for all $k$ and $s$, 
    \[x^k\RPref (\frac{\sum_{k'}\sum_{s'} h^k_s (k',s')x^{k'}_{s'} }{\mu^*_s})_s\] for all $k$ implies that it is not the case that 
    \[x^k\RPrefS (\sum_{k'}\sum_{s'} h^k_s (k',s')\frac{x^{k'}_{s'}}{\mu^*_s})_s\] for some $k$.
    \item\label{it:AIOEU} There exists $U^k_s\geq 0$ and $\la^k>0$  such that \[U^k_s \leq U^{k'}_{s'} + \la^{k'} \frac{p^{k'}_{s'}s}{\pi_{s'}} (x^k_s - x^{k'}_{s'}) \]
    \item\label{it:RATOEU} The data is $\U_{\mathit{EU}}$-rational.
\end{enumerate}
\end{theorem}

Statement~\eqref{it:AIOEU} provides Afriat inequalities for the problem at hand. The equivalence between~\eqref{it:AIOEU} and~\eqref{it:RATOEU} is due to \cite{varian1983} and \cite{green1986}.  The rest of Theorem~\ref{thm:chrisEU} is due to \cite{chambers2016EUtest}. 

The property in Statement~\eqref{it:Chris1} is a revealed-preference axiom similar to Fishburn's \citep{fishburn1975separation}, but in the consumption theory context. The axiom is a simple consequence of $\U_{\mathit{EU}}$-rationality: If each $x^k$ is revealed preferred to $z^k$, and we take a convex combination of the $x$'s and the $z$'s that yield identical lotteries, then the convex combination cannot place positive weight on any $x^k$ that is strictly revealed preferred to the corresponding $z^k$. Put differently, if I have two ways of composing the same (reduced) lottery, one with support in the $x^k$s and the other with support in the $z^k$s, the first collection of lotteries cannot all be better than the second collection, with some of them being strictly better. It is easy to see that this property~\eqref{it:Chris1} is implied by EU. What is surprising here is that it is also sufficient for $\U_{\mathit{EU}}$-rationality. 

It may be of interest to see that the property in Statement~\eqref{it:Chris1} implies GARP. Suppose given a sequence of observations $x^{k_i}$, $1\leq M$ ($M\geq 2$), with $x^{k_i}\RPref x^{k_{i+1}}$ for $1\leq i\leq M-1$, and $x^{k_M}\RPrefS x^{k_{1}}$. Then letting $z^{k_i}=z^{k_{i+1}}$ (using summation mod $M$), and a uniform distribution over the $x^{k_i}$ and $z^{k_i}$ gives a violation of Statement~\eqref{it:Chris1}.

The property in Statement~\eqref{it:Chris2} results from applying
linear programming duality to the system of Afriat inequalities
in~\eqref{it:AIOEU}. Since $\mu^*_s = \sum_{k'}\sum_{s'} h^k_s (k',s')
$, the function $h^k_s$ can be thought of as providing mean preserving
spreads. Then \eqref{it:Chris1} can be reduced to \eqref{it:Chris1}:
see \cite{chambers2016EUtest} for details. The equivalence
between~\eqref{it:Chris2} and~\eqref{it:AIOEU} is a matter of Farkas
lemma (see \cite{chambers2016revealed} for an exposition of this
methodology as it applies to revealed preference theory).

\subsection{Intertemporal choice}

I now turn to intertemporal choice with multiple physical goods, and present a result due to Martin Browning. The model is restricted to a dataset with $K=1$. That is, with a single observation. We call such datasets {\em survey data} because they are common in cross-sectional consumption surveys. 

A survey dataset $(x,p) = ((x_t)_{t=1}^T,(p_t)_{t=1}^T)$  satisfies cyclic monotonicity if, for any sequence of observations, $(x_{t_{i}}, p_{t_{i}})_{i=1}^M$
\[ 
\sum_{i=1}^M p_{t_i}\cdot (x_{t_{i+1}} - x_{t_{i}})\leq 0,
\] where we are employing addition mod $M$ in the calculation of subindeces of $t$. 

Let $\U_{\mathit{NDU}}$ be the set of all utility functions $U$ such that
\[ U(x_1,\ldots,x_T) = \sum_{t=1}^T u(x_t),
\] where $u:\Re^L_+\rightarrow\Re$ is continuous and concave.

The next result is due to \cite{browning1989nonparametric}.

\begin{theorem}\label{thm:browning}
A survey dataset (a dataset with $K=1$) is $\U_{\mathit{NDU}}$ rational if and only if it satisfies cyclic monotonicity. 
\end{theorem}

The idea behind Theorem~\ref{thm:browning} is simple. Suppose, for the sake of exposition, that a rationalizing $u$ is differentiable. Then the first-order condition that a $\U_{\mathit{NDU}}$ rational dataset must satisfy is \[ 
D u(x_t) = \la p_t, 1\leq t\leq T.
\] This means that $1/\la p_t$ must constitute the values of the gradient of a concave funciton. It is well known that cyclic monotonicity characterizes the gradients of concave functions (see \cite{rockafellar1997convex}).

\cite{crawford2010habits} generalizes the ideas in \cite{browning1989nonparametric} to study a model with exponential discount and habit formation. See also \cite{demuynck2013ll} for a revealed preference analysis of habit-forming durable goods.

\section{Conclusion}\label{sec:conclusion}

Revealed preference theory has traditionally explored the empirical content of the theory of optimizing behavior. An emergent literature has extended the theory to cover specific functional forms that are important in specific economic environments. In this survey, I have reviewed the recent literature  focused on the economic environments of risk, uncertainty, and intertemporal choice. The last five years have seen significant advances in our understanding of the empirical content of some of the most widely used models of choice for such environments.

These theoretical advances are mirrored in the emergence of novel experimental designs for choice under economic budget sets. The resulting  experimental datasets provide an excellent opportunity for testing important economic theories under controlled conditions and weak assumptions. My survey has focused on theoretical results, but several recent papers have used the revealed-preference tests developed in the theoretical work, and applied them to experimental data. 

\bibliographystyle{ar-style1}
\bibliography{chapterRP}

\end{document}